\documentclass{iopart}

\usepackage{graphicx} 
\usepackage{iopams}

\newcommand{\be}{\begin{equation}} \newcommand{\ee}{\end{equation}}
\newcommand{\bea}{\begin{eqnarray}} \newcommand{\eea}{\end{eqnarray}}

\def\half{{1 \over 2}}  \def\pa{{\partial}}

\def\({\left(} \def\){\right)} \def\[{\left[} \def\]{\right]}
\newcommand{\ltsim}{\protect\raisebox{-0.5ex}{$\:\stackrel{\textstyle
      <}{\sim}\:$}}

\def\al{\alpha} \def\bt{\beta} \def\kp{\kappa} 
 \def\lam{\lambda}  \def\eps{\epsilon}
\def\sig{\sigma}

\begin{document}
\title{On critical collapse of gravitational waves} \author{Evgeny
  Sorkin}
\address{Max Planck Institute for Gravitational Physics (Albert Einstein Institute) \\
  Am Muehlenberg 1, Potsdam, 14476, Germany}
\ead{Evgeny.Sorkin@aei.mpg.de}
\begin{abstract}
  An axisymmetric collapse of non-rotating gravitational waves is
  numerically investigated in the subcritical regime where no black
  holes form but where curvature attains a maximum and decreases,
  following the dispersion of the initial wave packet. We focus on a
  curvature invariant with dimensions of length, and find that near the
  threshold for black hole formation it reaches a maximum along
  concentric rings of finite radius around the axis. In this regime
  the maximal value of the invariant exhibits a power-law scaling 
  with the approximate exponent 0.38, as
  a function of a parametric distance from the threshold.  In addition,
  the variation of the curvature in the critical limit is accompanied
  by increasing amount of echos, with nearly equal temporal and
  spatial periods.  The scaling and the echoing patterns, and the
  corresponding constants are independent of the initial data and
  coordinate choices.
\end{abstract}
\pacs{04.25.D-,04.25.dc,04.20.-q}
\section{Introduction}
\label{sec_intro}
Universality, scaling and self-similarity found in critical
gravitational collapse is one of the most fascinating phenomena
associated with gravitational interactions. First discovered
numerically by Choptuik \cite{Choptuik} in spherically-symmetric
collapse of massless scalar field, this distinctive behaviour was later
observed in other systems, including those with various matter
contents and equations of state, diverse spacetime dimensions
etc. However, while a great deal of literature has emerged on critical
phenomena in spherical symmetry, only a limited number of
non-perturbative studies exists in less symmetric settings, see
\cite{Gundlach_review} for a review.

Perhaps the simplest non-spherical system is a pure axisymmetric
gravitational wave, collapsing under its own gravity. Abrahams and
Evans \cite{AbrahamsEvans} found that the mass of black holes, forming
in the evolution of sufficiently strong initial waves, exhibits
scaling of the form $M_{bh} \propto (a-a_*)^\bt $ in the limit when
the strength parameter $a$ tends to $a_*$, the threshold for black
hole formation, and determined the exponent of the power-law to be
$\bt \simeq 0.35-0.38$.  They have also given less conclusive evidence
of periodic echoing of the near critical solutions.  Surprisingly,
these results proved difficult to reproduce; in fact no other
successful simulation of the axisymmetric vacuum collapse has been
reported to date (see e.g. \cite{Seidel_etal} for a failed
attempt). However, in this paper we present new results obtained with
the aid of our recent harmonic code \cite{ES_2+1+n}.

We focus on subcritical collapse of axisymmetric non-rotating Brill
waves during which black holes do not form, but where curvature grows
to reach a maximum and subsequently diminishes, following the
dispersion of the initial wave.  Perturbative studies of the critical
solutions \cite{Gundlach_review,Koike_etal} suggest that the power-law
scaling of characteristic quantities near the critical point should
occur on both sides of the black hole formation threshold, regardless
of appearance of horizons.  While this was confirmed in numerical
experiments in spherical symmetry, see e.g
\cite{Garfinkle_sub,SorkinOren}, it is an open question whether same
is true in other situations as well. Here we demonstrate that in the
axisymmetric subcritical collapse, a curvature invariant with
dimensions of length follows a power-law with the exponent,
$\bt=0.385\pm0.015$, similar to that found by Abrahams and Evans in
supercritical case.  Additionally, we find that the solutions develop
increasingly large numbers of echos as the critical limit is
approached. Our current resolution allows observation of up to three
echos around the time instant where curvature is maximal.  We measure
that, for example, the Riemann curvature invariant oscillates in time
with the (logarithmic) period of $\Delta_\tau \simeq 1.1$, and that
the logarithm of the invariant changes on each echo by nearly same
amount $\Delta_r \sim \Delta_\tau\simeq1.1$.

We verify that the scaling and echoing constants are essentially
independent of initial data and specific coordinate conditions used to
calculate the solutions. In contrast to spherically symmetric
collapse, where the greatest curvature is always at the origin, the
evolution of the axisymmetric waves is more complicated and the
spacetime location of the maximum depends strongly on the geometry of
the initial data. We evolved series of subcritical initial data where curvature
attained a maximum along equatorial rings of various radii centered at
the axis. Besides, we found that in supercritical evolutions of the
same data an apparent horizon forms, engulfing the ring-shaped locus of the maximal curvature. 
This indicates that the critical solutions found in the
Brill-wave evolutions are different from ones calculated by Abrahams \& Evans, in whose case the
maximal curvature has always occurred at the origin, and the
black holes tend to be arbitrary small in the critical
limit. Strikingly,  despite these differences the near critical scaling and echoing patterns are
similar, and the scaling exponents are comparable.

In the next section we briefly describe the initial value problem for
constructing the axisymmetric vacuum asymptotically flat spacetimes
without angular momentum; the details of the equations, gauge
conditions and our numerical code are found in \cite{ES_2+1+n}.
Section \ref{sec_results} is devoted to the results and numerical
tests. We summarize our findings, discuss limitations of the current
method and outline perspectives in the concluding section
\ref{sec_discussion}.

\section{The equations and a method of their solution}
\label{sec_setup}
We are interested in solving the vacuum Einstein equations
\be
\label{Rmunu}
R_{\mu\nu}=0, \ee
where $R_{\mu\nu}$ is the Ricci tensor. We consider axisymmetric
asymptotically flat spacetimes without angular momentum, and assume
they can be foliated by a family of spacelike hypersurfaces, starting
with the initial surface at $t=0$, where the spatial metric and its
normal derivatives are chosen to satisfy the constraints
(Gauss-Codazzi equations).

The most general metric adapted to the symmetries of the problem can
be written using the cylindrical coordinates
\be
\label{ds2}
ds^2=g_{ab}\, dx^a\,dx^b + r^2 e^{2\,S} d\phi^2, \ee
where the seven metric functions---$g_{ab}$, $a,b=0,1,2$ and
$S$---depend only on $t,r$ and $z$.  \footnote{While Greek indices
  range over $t,r,z,\phi=0,1,2,3$, Latin indices run over $0,1,2$.}
In order to solve the field equations (\ref{Rmunu}) we employ the
Generalized Harmonic (GH) formalism
\cite{Friedrich_GH,Garfinkle_GH,Pretorius_GH}, adopted to the axial
symmetry in \cite{ES_2+1+n}. To this end we define the GH constraint,
\be
\label{C_GH}
C^a \equiv -\Box x^a +H^a =-\Gamma^a_{\al\bt} g^{\al\bt} +H^a= 0, \ee
where $\Gamma^a_{\al\bt}$ are Christoffel symbols, and the ``source
functions'' $H^a=H^a(x,g)$ depend on coordinates and the metric (but
do not on the metrics's derivatives) and are arbitrary otherwise. We
than modify the Einstein equations:
\be
\label{GH_eqs}
R_{\mu\nu} - C_{(\mu;\nu)} =0, \ee
that now become a set of quasilinear wave equations for the metric
components of the form $g^{\al\bt} g_{\mu\nu,\al\bt} +\dots = 0$,
where ellipses designate terms that may contain the metric, the source
functions and their derivatives.

Fixing the coordinate freedom in the GH language amounts to specifying
the source functions, and we choose those by requiring that the
spatial coordinates satisfy damped wave equations, while the time
coordinate remains well behaved while the lapse satisfies
a damped wave equation \cite{DWG,ES_2+1+n}. A particular example
of these conditions \cite{DWG}, that we use here, can be written in
terms of the kinematic ADM variables as
\be
\label{H_DWG}
H_a^{DW}=2\,\mu_1\,\log\(\frac{\gamma^{1/2}}{\al}\)\,n_a-2\,\mu_2\,\al^{-1}\,\gamma_{ai}\,\bt^{i},
\ee
where $n_\mu=(-g^{00})^{-1/2} \pa_\mu\,t$ is the unit normal to the
spatial hypersurfaces of constant time, $\al$ is the lapse, $\bt^i$ is
the shift, $\gamma_{ab}=g_{ab}+n_a n_b$ is the spatial metric,
$\gamma=(g_{11}\,g_{22}-g_{12}^2)\,\exp(S)$, and $\mu_1$ and $\mu_2$
are parameters.

The initial data is given at $t=0$, where we choose the initial
spatial metric to be in the form of the Brill-wave \cite{Brill}
\be
\label{Brill_id}
ds^2_3=\psi^4(r,z)\[e^{2\,r\,B(r,z)}\( dr^2+dz^2\)+ r^2\, d\phi^2\],
\ee
with
\be
\label{B}
B(r,z)\equiv a \, r
\exp\(-\frac{r^2}{\sig_r^2}-\frac{z^2}{\sig_z^2}\), \ee
where $\sig_r, \sig_z$ and $a$ are parameters.

We further assume time-symmetry, in which case the momentum constraint
identically vanishes at $t=0$, while the Hamiltonian constraint
becomes the elliptic equation for $\psi$
\be
\label{Eq_psi}
\(\pa_r^2 + \frac{1}{r}\pa_r+\pa_z^2\)\psi=-\frac{1}{4} \psi\, r\,
\(\pa_r^2 +\frac{2}{r} \pa_r +\pa_z^2\)B, \ee
which is solved subject to regularity conditions at the axis,
equatorial reflection symmetry, and asymptotic flatness boundary
conditions:
\be
\label{bc_psi}
\pa_z \psi(r,0)=0, ~~ \pa_r \psi(0,z)=0, ~~
\psi(r,\infty)=\psi(\infty,z)=1.  \ee
We assume initially harmonic coordinates, $H^a=0$, and choose the
initial lapse $\al(t=0,r,z)\equiv g_{00}^{1/2}(t=0,r,z)=1$.

Having specified the initial data we integrate the equations
\eref{GH_eqs} forward in time, imposing asymptotic flatness and
regularity at the axis, $r=0$. For simplicity, we restrict attention
to the spacetimes having equatorial reflection symmetry.  The
highlights of our finite-differencing approximation (FDA) numerical
code \cite{ES_2+1+n} that we employ to solve the equations include:

\begin{itemize}
\item An introduction of a new variable that facilitates axis
  regularization.  While elementary flatness at the axis implies that
  each metric component has either to vanish or to have vanishing
  normal derivative at that axis, requiring absence of a conical
  singularity at $r=0$ results in the additional condition:
  $g_{11}(t,0,z)=\exp[2\,S(t,0,z)] $. Therefore, at $r=0$ we
  essentially have three conditions on the two fields $S$ and
  $g_{11}$.  While in the continuum, and given regular initial data,
  the evolution equations will preserve regularity, in a FDA numerical
  code this will be true only up to discretization errors. Our
  experience shows that the number of boundary conditions should be
  equal to the number of evolved variables in order to avoid
  regularity problems and divergences of a numerical
  implementation. We deal with this regularity issue by defining a new
  variable
  \be
  \label{lambda}
  \lambda \equiv \frac{g_{11} -e^{2\, S}}{r} , \ee
  that behaves as $\lam \sim O(r)$ at the axis, and use it in the
  evolution equations instead of $S$. This eliminates the
  overconstraining and completely regularizes the equations.
  Crucially, the hyperbolicity of the GH system is not affected by the
  change of variables.

\item Constraint damping. The constraint\footnote{It can be shown that
    the standart Hamiltonian and momentum constraints are equivalent
    to the GH constraints \cite{Lindblom_etal1}.}  equations,
  $C_\mu=0$, are not solved in the free evolution schemes like ours,
  except at the initial hypersurface.  While one can show that in the
  continuum the constraints are satisfied at all times, in FDA codes
  small initial violations tend to grow and destroy convergence. A
  method that we use to damp constraints violations consists of adding
  to the equations (\ref{GH_eqs}) the term of the form
  \cite{Gundlach_CD,Pretorius_BH1},
  \be \label{cdmp} Z_{\mu\nu} \equiv \kappa \( n_{(\mu} C_{\nu)}
  -\half g_{\mu\nu} \, n^\bt \, C_\bt \), \ee
  where $\kappa$ is a parameter. We note that $ Z_{\mu\nu}$ contains
  only first derivatives of the metric and hence does not affect the
  principal (hyperbolic) part of the equations.

\item A spatial compactification is introduced in both spatial
  directions by transforming to the new coordinates ${\bar x} =
  x/(1+x), {\bar x} \in [0,1], x\in [0,\infty)$, where $x$ stands for
  either $r$ or $z$.  The advantage of this scheme is that asymptotic
  flatness conditions $g_{\mu\nu}=\eta^{Mink}_{\mu\nu}$ at the spatial
  infinity are exact.

\item We use Kreiss-Oliger-type dissipation in order to remove high
  frequency discretization noise. \footnote{ i.e. the noise with
    frequency of order of the inverse of the mesh-spacing} An
  additional role of the dissipation is to effectively attenuate the
  unphysical back reflections from the outer boundaries, resulting
  from the loss of numerical resolution there. This allows using
  compactification meaningfully \cite{Pretorius_GH,SorkinChoptuik}.

\end{itemize}

In order to characterize the spacetimes that we construct, we use the
Brill mass \cite{Brill}, computed at the initial time-slice,
\be
\label{M_B}
M=\int \[\(\pa_r \log \psi\)^2 + \(\pa_z \log \psi\)^2\] \,r\,dr\,dz,
\ee which---we verify---coincides with the ADM mass.
For the purpose of quantifying the strength of the gravitational field
we calculate the Riemann curvature invariant having dimension of
inverse length,
\be
\label{Riemann2}
I \equiv (R_{\al\bt\mu\nu}R^{\al\bt\mu\nu})^{1/4} \ee
at various locations, and in certain experiments we follow its
evolution in the proper time at that location $(r,z)$,
\be
\label{tau}
\tau(t,r,z) \equiv \int^t_0 \al(t',r,z) dt'.  \ee
We also use the circumferential radius \be
\label{rho}
\rho \equiv r\, e^S.  \ee
\section{Results}
\label{sec_results}
The initial data (\ref{Brill_id},\ref{B}) are characterized by the
amplitude $a$ and the ``shape'' parameters $\sig_r$ and $\sig_z$,
which define the mass of the data and their ``strength'', namely the
tendency to collapse and form a black hole.  For a given amplitude and
fixed $\sig_r+\sig_z=const$, the data with larger $\sig_z/\sig_r$ are
stronger (see also \cite{Teukolsky_BW,Garfinkle_BW}). In addition, by
varying the shape parameters at fixed gauge, we can control the
spacetime locations where curvature evolves to a maximum or where an
apparent horizon first forms.  In our experiments we use several sets
of $\sig_r$ and $\sig_z$, and adjust the strength of the initial wave
by tuning its amplitude.

The initial data are numerically evolved forward in time.  We use
grids with similar mesh-sizes in both spatial dimensions $h_r=h_z=h$,
and time-steps of $h_t=0.04\,h$ and $h_t=0.05\,h$.  Usually our fixed
grids consist of 200, 250, 300 or 400 points, uniformly covering the
compactified spatial directions.  We also experiment with adaptive
mesh refinement (AMR), provided by the PAMR/AMRD software \cite{pamr}.
In this case we use two or four refinement levels, and the base mesh
with the resolution of $h=1/128$.  The Kreiss-Oliger dissipation
parameter is typically $\eps_{KO}=0.5-0.85$, with larger values used
on finer grids and stronger initial data; and the constraint damping
parameter in (\ref{cdmp}) is $\kappa=1.4-1.7$.  The gauge fixing
parameters (\ref{H_DWG}) in the ranges $\mu_1 \simeq 0.1-0.3$ and
$\mu_2 \simeq 0.9-1.2$, usually gave stable, sufficiently long
evolutions.


\begin{table}
  \begin{indented}
  \item[] \hspace{-0.8cm}
    \begin{tabular}{@{}ccccccccccc}
      \br
      $\sig_r, ~\sig_z$ &$\mu_1,~\mu_2$ & $h$ &$h_t/h$ & $a_*$ &$M_*$ & $\rho_*$ & $\tau_*$ \\
      \br
      1.0,~1.0 &0.1,~1.1 & 1/200 & 0.05& $5.985\pm 0.005$ &0.969 & 0.2 & 1.5  \\
      1.0,~1.0 &0.12,~1.17 & 1/300 &0.05 &$6.20021\pm 0.00001$ &1.04 &0.15 & 1.46   \\
      1.0,~1.0 &0.2,~1.0 &1/400 & 0.04&$6.273\pm 0.001$ &1.06&0.2 & 1.45   \\
      0.9,~1.3& 0.3,~0.9 & 1/300 &0.04 & $ 7.3079082\pm 0.0000002$& 1.12 & 0.23  & 1.59 &    \\
      0.9,~1.3& 0.3,~0.9  & 1/128,~ 2l &0.05 &$7.246067\pm 0.000002$ &  1.1  & 0.23 & 1.64   \\
      0.9,~1.3& 0.3,~0.9  & 1/128,~ 4l &0.05 &$6.9401\pm 0.0002$ &  1.0  & 0 & 1.51   \\
      0.8,~1.3& 0.2,~1.1  & 1/300 &0.05 &$6.786\pm 0.004$ &  0.607  & 0.1 & 3   \\
      0.7,~1.5& 0.2,~1.1 & 1/300 &0.04 &$8.21\pm 0.01$ &  0.593 & 0.0 & 3.2   \\
      \br
    \end{tabular}
  \end{indented}
  \hspace{-1cm}
  \caption{\label{tab_sigs}
    The parameters of the initial data $\sigma_r, \sigma_z$, as well as grid spacing $h$ and gauge parameters $\mu_1, \mu_2$
    determine the threshold amplitude $a_*$ whose upper margin corresponds to formation
    of a black hole, and whose lower margin corresponds to a regular spacetime. Given this set of initial parameters,
    this further determines the total mass $M_*$ and the ``accumulation locus'', whose position and time of occurrence
    is given by the radial position $\rho_*$ and proper time $\tau_*$. The radial position $\rho_*$ is measured in terms of
    the circumferential radius (\ref{rho}), and the proper time $\tau_*$ at that location is measured in units of the total mass.
    The parameters $2l$ and $4l$ indicate that 2 and 4 AMR levels were used, respectively. All other simulations are unigrid. }
\end{table}


The system is weakly gravitating for small amplitudes, in which cases
the initial wave packet ultimately disperses to infinity. However, for
amplitudes above certain threshold, $a_*$, the wave collapses to form
a black hole, signaled by an apparent horizon.  In subcritical
spacetimes we can define the ``accumulation locus'' where curvature
attains a global maximum before decaying. In our coordinates
(\ref{H_DWG}), and for our initial data (where the ratio of  
$\sig$'s never exceeds five) the position of the maxima
$(t_*,r_*,z_*)$ is always along the equator $z_*=0$.

The threshold amplitude for black hole formation, $a_*$, depends on
the initial data, controlled by $\sigma_r,\sigma_z$, and gauge
parameters $\mu_1,\mu_2$, and the resolution, $h$. Table
\ref{tab_sigs} records critical amplitudes, $a_*$, masses, $M_*$ and
the spacetime positions, $\rho_*, \tau_*$, of the accumulation locus
in the strongest, $a\simeq a_*$, initial data evolutions, for a few sets
which we have calculated.  In contrast to spherically-symmetric
collapse, where accumulation locus is solely at the origin, in axial
symmetry this is not always the case. For instance, the critical
amplitude for the initial data with  $\sig_r=\sig_z=1$, determined in the unigrid simulations with 
$h=1/300$  is $a_*=6.20021$. The spacetime position of the accumulation
depend on the amplitude such that for  $a\ltsim 0.99
a_*$ the accumulation loci are at the origin, and for larger
amplitudes they shift to be along the rings of radii $\rho_*
\simeq 0.2$. The time of occurrence of the maxima converges to
$\tau_*\simeq 1.46\, M_*$ from above in the limit $a\rightarrow
a_*$. While qualitatively similar behaviour is observed for most of the initial
data families listed in Table \ref{tab_sigs}, the initial data defined by
$\sig_r=0.7, \sig_z=1.5$ has the accumulation loci at the origin all
the way to the strongest subcritical amplitude of  $a=8.20$. However,
since in this case we have succeeded to compute $a_*$ only with a
modest accuracy  of 1 part in 820, a possibility remains that closer to the threshold the
accumulation loci will become ring-shaped.  For this set the time of
the accumulations 
converges to $\tau_*\simeq 3.2\, M_*$ in the limit $a\rightarrow a_*$, and the mass of the near
critical solutions, $M_* \simeq 0.593$, is about one half of that found in the
$\sig_r=0.9, \sig_z=1.3$ case. 

As described next, there is a power-law scaling of the maximal curvature in the
limit $a/a_*\rightarrow 1$ for all families of the initial data listed
in Table \ref{tab_sigs}.  In most cases the scaling shows up at
relatively large values of $a_*/a-1 \sim 10^{-3}$, for all resolutions
better than $h=1/200$. \footnote{For comparison, in scalar field
  collapse the signatures of near-critical scaling do not appear
  before $a_*/a-1 \lesssim 10^{-8}$.} However, it turns out that the
data calculated in fixed-mesh simulations with $h \gtrsim 1/250$ is
too noisy and dependent on the details of numerics, to provide a
reliable estimate of the scaling exponent. 

\begin{figure}[t!]
  \centering 
  \includegraphics[width=7cm]{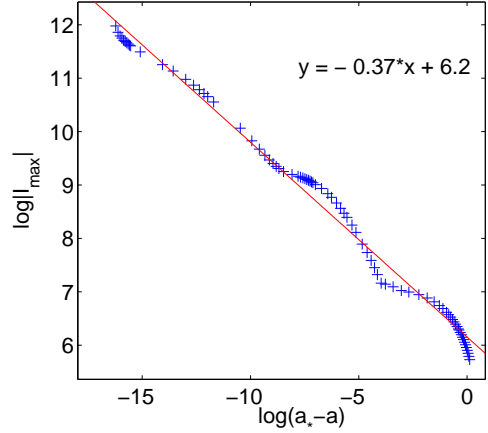}
  \caption[]{ A logarithm of the maximal Riemann invariant
    (\ref{Riemann2}) as a function of the distance from the critical
    amplitude, $a_*-a$, in the simulations with $\sig_r=0.9, ~
    \sig_z=1.3$, and the fixed resolution $h=1/300$. The critical
    amplitude in this case is $ a_*=7.3079082$, and the maximal
    curvature is $|I_{max}|\sim 10^{5}$ in the units of the total
    mass.  The linear fit to the data (solid line) has the slope
    $\bt\simeq-0.37$. Notice the (quasi-) periodic ``wiggle'' of the
    data points about the straight line, which we interpret to signal periodic
    self-similarity of the critical solution.}
  \label{fig_loglogIa_obt0913}
\end{figure}
\begin{figure}[t!]
  \centering 
  \includegraphics[width=7cm]{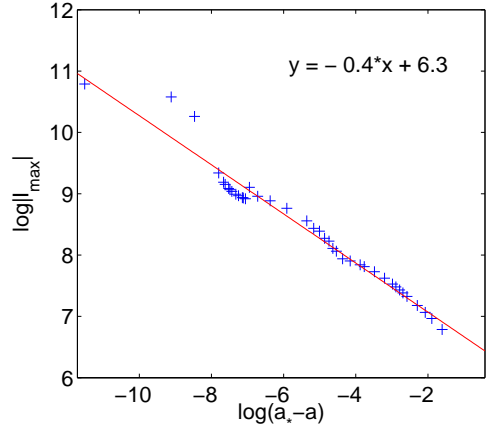}
  \caption[]{A plot similar to Fig. \ref{fig_loglogIa_obt0913}, but
    obtained with different parameters: $\sig_r=\sig_z=1$ and the
    resolution $h=1/300$.  In this case the critical amplitude is
    $a_*=6.20021$.  Remarkably, the slopes of the linear fits in both
    figures agree to within $8 \%$. }
  \label{fig_loglogIa_kp17}
\end{figure}
The scaling can be envisaged by plotting the maximal value of the
Riemann curvature invariant (\ref{Riemann2}) as a function of the
parametric distance from the critical amplitude, $a_*-a$; this is
shown in Fig. \ref{fig_loglogIa_obt0913}. Each point here represents
the global maximum $|I_{max}|$ computed during evolutions defined by
$\sig_r=0.9$ and $\sig_z=1.3$, and the numerical parameters: $h=1/300,
h_t/h=0.04, \mu_1=0.3, \mu_2=0.9, \kp=1.7, \eps=0.6$. The solid line
represents the least-squares linear fit to the data. The slope of the
line, $\bt \simeq -0.37$, is in agreement with the exponent of the
black-holes' mass scaling, \footnote{Note that our exponent is
  negative since the dimensions of $I$ are inverse length, while black
  hole mass computed in \cite{AbrahamsEvans} has dimensions of length.}  found in supercritical
collapse by Abrahams and Evans \cite{AbrahamsEvans}.  The data
depicted in Fig. \ref{fig_loglogIa_kp17} were obtained with a
differently shaped initial wave, $\sig_r=\sig_z=1$, and the parameters
$h=1/300, h_t/h=0.05, \mu_1=0.12, \mu_2=1.17, \kp=1.7, \eps=0.8$.  The
threshold amplitude in this case is found with somewhat lesser
accuracy, $a_*=6.20021 \pm0.00001$.  However, the data is still fitted
well with a straight line whose slope, $\bt \simeq-0.4$, coincides
with the exponent in Fig. \ref{fig_loglogIa_obt0913} to within $8 \%$.

\begin{figure}[t!]
  \centering 
  \includegraphics[width=12.5cm]{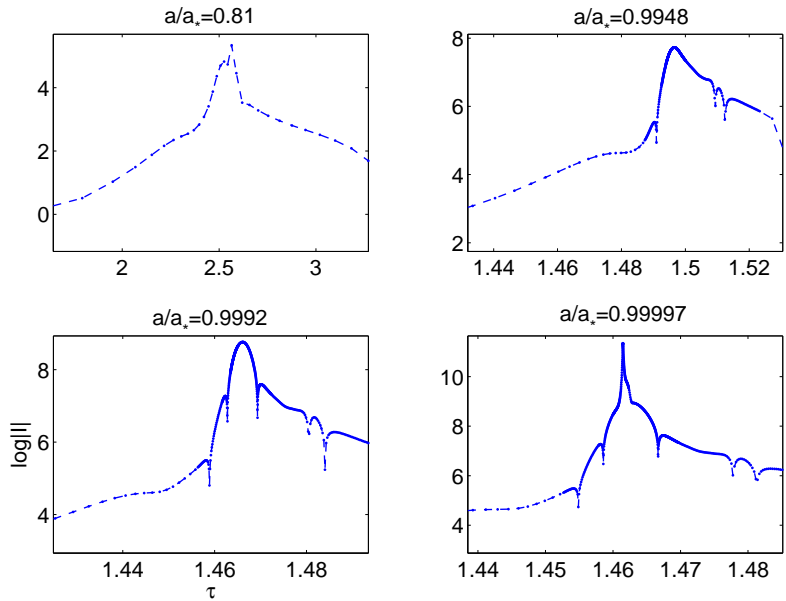}
  \caption[]{The dynamics of the logarithm of the Riemann invariant,
    $I$, as a function of the proper time $\tau(r_*)$ for several
    values of $a/a_*$, obtained in the evolution of initial data
    defined by $\sigma_r=0.9, \sigma_z=1.3$, and the fixed resolution
    $h=1/300$.  The variation of $I$ toward the accumulation,
    $\tau_*$, and away from it is accompanied by oscillations, whose
    number grows in the limit $a \rightarrow a_*$. The double dip in
    top right panel at about $\tau\simeq 1.51$ and in bottom panels
    near $\tau\simeq 1.48$ is a result of the interference between the
    main and a secondary reflection off the axis.}
  \label{fig_peaks}
\end{figure}

It is remarkable that despite the evolutions of the initial waves
shown in Figs. \ref{fig_loglogIa_obt0913} and \ref{fig_loglogIa_kp17}
are dramatically different, the maximal curvatures in both cases
follow a power-law with similar exponents.  We verify that the same
scaling again appears in simulations with other shape parameters and
in all cases that resulting exponent is consistently in the range
$\bt\simeq 0.37-0.4$. In addition, the scaling exponent within these
bounds
 when we use coordinate conditions with different choices of
$\mu$'s in (\ref{H_DWG}) (see e.g. Figs. \ref{fig_loglogIa_obt0913}
and \ref{fig_loglogIa_kp17}).  While this does not test the rigidity
of $\bt$ with respect to all possible coordinate conditions, this
demonstrates relative consistency of the exponent within the large
family of the gauges (\ref{H_DWG}). We conclude that in the critical
limit the maximal curvature predominantly scales as $|I_{max}| \propto
(a_*-a)^{-\bt}$, with $\bt=0.385\pm0.015$, where the errorbars
represent the deviation from the average value computed over all
initial data sets that we have evolved.\footnote{The slope obtained
  for each initial data set carries individual fitting
  errors. However, these are typically smaller than the fluctuations
  around the average $\beta$ computed over all data sets.}

The distribution of data points in Figs. \ref{fig_loglogIa_obt0913}
and \ref{fig_loglogIa_kp17} has a striking property, namely the data
``wiggle'' about the linear fit. We note that similar wiggle was also
observed in near critical collapse of scalar field. In that case it
was attributed \cite{GHP} to the periodic self-similarity found in
that system, where the critical solution, $Z_*$, repeats on itself
after a discrete period $\Delta$:
$Z_*(\tau,r)=Z_*(\tau\,e^\Delta,r\,e^\Delta)$.  Besides, \cite{GHP}
found that the period of the wiggle is $\Delta/(2\,\bt)$, and thus
may, in principle, allow calculating the self-similarity scale
$\Delta$ by measuring the slope and the period of the wiggle in a plot
like ours Figs. \ref{fig_loglogIa_obt0913} and
\ref{fig_loglogIa_kp17}. We believe that the quasi-periodic fluctuations of
the points about the linear fit in these figures do signal discrete
self-similarity, however our current data are insufficiently accurate and have
too short a span to provide more quantitative estimate of the wiggle
period, beyond a very rough value of anything between two and four.

Independent, and more direct signatures of discrete self-similarity
are obtained by examining the behaviour of the curvature when
$a \rightarrow a_*$.  It turns out that in this limit, in addition to
that $I$ attains increasingly larger maxima, the temporal variation of
$I$ is also accompanied by increasing amount of oscillations.  This is
illustrated in Fig. \ref{fig_peaks}, which shows the variation of $I$
as a function of the proper time, calculated at the accumulation loci, for a
sequence of $a$'s. Figure shows that the amount of
fluctuations---indicated by the peaks or inflection points---grows from
one to three in the limit $a/a_* \rightarrow 1$ on both sides of the
accumulation locus. Such an oscillatory behaviour is again reminiscent
of the ``echoing'' in critical spherical collapse of scalar field (see
e.g. Fig. 5 in \cite{SorkinOren} and Fig. 7 in \cite{HamadeStewart})
and we interpret it as evidence of periodic self-similarity in our
system as well.

\begin{figure}
  \centering 
  \includegraphics[width=13cm]{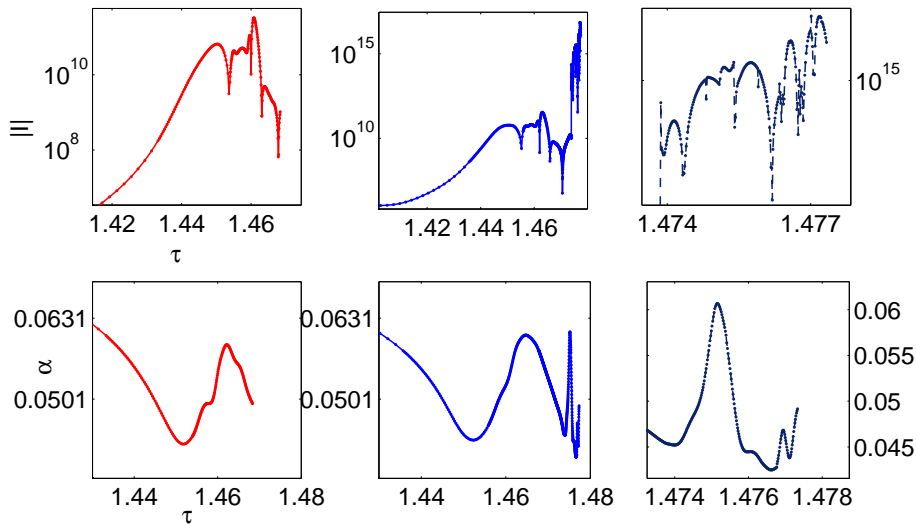}
  \caption[]{The echoing pattern obtained in the evolution of the
    initial data sets with $\sig_r=0.9, ~\sig_z=1.3$ and
    $a=6.940$. Left panels show the low resolution runs that use 2
    levels of AMR, with the base $h=1/128$, other panels were obtained
    using 4 AMR levels with the same base; the rightmost panels is the
    zooming of the late time behavior shown in middle panels.  While
    the lower resolution runs diverge around $\tau\simeq 1.47$, the
    higher resolution runs extend beyond that, allowing to calculate
    additional echos. Notice that $I$ has sharper and easier to
    identify features than $\alpha$. }
  \label{fig_peaks_obt}
\end{figure}

Like the power-law scaling of the maximal curvature, the echoing of
our solutions in the near critical limit is independent of specific
gauges or particular initial data sets.  This is demonstrated in
Fig. \ref{fig_peaks_obt}, which depicts temporal evolutions of the
Riemann invariant and the lapse function found in simulations of the
initial data characterized by $\sig_r=0.9,~\sig_z=1.3$, the gauge
constants $\mu_1=0.3,~ \mu_2=0.9$, and the amplitude $a=6.940$.  The
functions in left panels were computed using 2 levels of AMR, and the
other panels were generated using 4 levels of AMR; in both cases the
base-level resolution is $h=1/128$. Figure shows that the dynamics
in this case involves more scatterings and interferences of the initial and secondary waves than e.g.
in $\sig_r=\sig_z$ runs, depicted  in Fig. \ref{fig_peaks}. 

In most cases higher resolution simulations run longer and
allow computing more oscillations.  Notice that the shapes of the
curves in left and middle panels in Fig. \ref{fig_peaks_obt} are
essentially identical until $\tau \sim 1.47 $.  However, while the lower
resolution runs diverge around that time due to formation of a
singularity, the higher resolution evolutions continue beyond
that, and develop additional echos that accumulate near $\tau_*\simeq1.477
M_*$, just before the numerics fails. The critical amplitude,
determined in the 4-level AMR simulations is $a_*=6.9401$, and the
accumulation loci occur are the origin.  While  we were
unable to stabilize the 4-level evolutions for amplitudes beyond
about $a\simeq 7$,  in lower-resolution, 2-level runs we find a different critical solution with the amplitude,
$a_*=7.246067$, where the accumulation locus lies at $\rho_*
\simeq 0.23$, see Table \ref{tab_sigs}.  The total masses
of the near critical spacetimes, the accumulation loci and such details of
evolutions as the amount of secondary scatterings and
interference's, reflected in the strong variability of the curvature
profile, and the total amount of gravitational radiation are
different in near-critical evolutions in the 2 and 4-level AMR
simulations. Nevertheless, the scaling and echoing constants
appear to be nearly identical.

\begin{figure}
  \centering 
  \includegraphics[width=7.5cm]{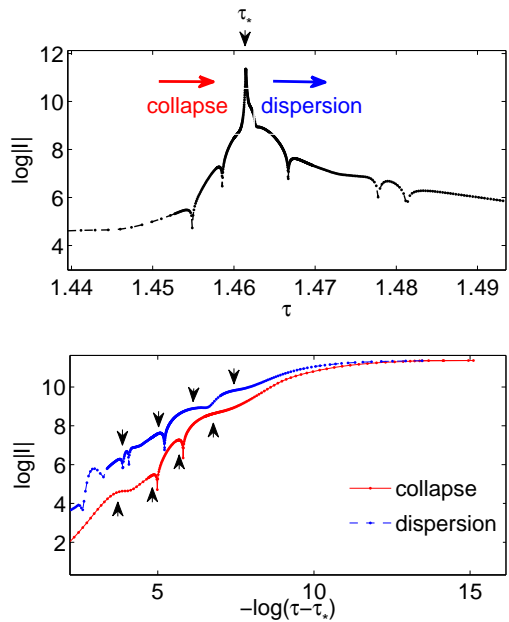}
  \caption[]{The typical temporal variation of the curvature invariant
    $I$ near the accumulation locus is oscillatory in time. Shown is
    the evolution of the initial data with $\sig_r=\sig_z=1$ and
    $a/a*=0.999998$.  On each oscillation $\log|I|$ varies by
    $\Delta_r\simeq 1.1 \pm 0.1$, which is close to the time period
    $\Delta_\tau\simeq0.95\pm0.15$ of the four oscillations around
    $\tau_*$.}
  \label{fig_echos}
\end{figure}
\begin{figure}
  \centering 
  \includegraphics[width=7.5cm]{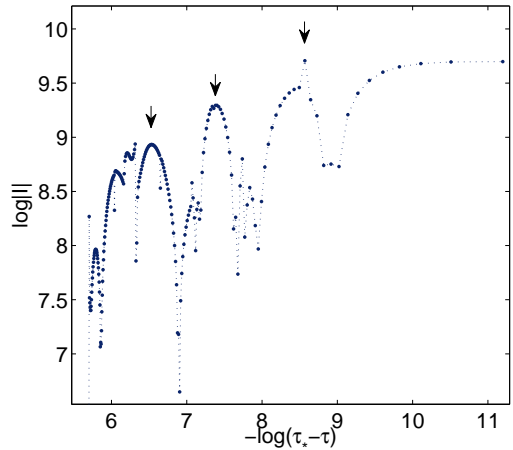}
  \caption[]{The variation of $I$ obtained in a 4-level AMR
    simulations initialized by $\sig_r=0.9,~\sig_z=1.3$, and
    $a/a*\simeq0.99997$.  The variation of $\log|I|$ on each
    oscillation (marked by arrows) is nearly equal to the temporal
    period of the echos, $\Delta_r\simeq\Delta_\tau \simeq 1.1 $.}
  \label{fig_echos_obt}
\end{figure}
In order to estimate the period of the echos we plot in
Fig. \ref{fig_echos} the temporal variation of $I$ computed in the
evolution of the initial data set having $\sig_r=\sig_z=1$ and
$a=6.2002$.  By measuring the distances between the peaks or
inflection points---marked by arrows in Fig. \ref{fig_echos}---we find
that the curvature fluctuates in time with the logarithmic period
$\Delta_\tau =0.95 \pm 0.15$ and that on each echo the logarithm of
$I$ varies by approximately $\Delta_r \simeq 1.1 \pm 0.1 $.  The
errorbars here represent the maximal deviation from the average values
of $\Delta_\tau$ and $\Delta_r$, measured in this figure. We note that
both periods agree within the error-bars.  A similar
Fig. \ref{fig_echos_obt} shows the dynamics of $I$ against
$\tau_*-\tau$, that was obtained in simulations with 4 levels of AMR,
$\sig_r=0.9, ~ \sig_z=1.3$, and $a=6.940$, shown in right panels in
Fig. \ref{fig_peaks_obt}. \footnote{The evolution of this initial data
  diverges soon after the accumulation at about $\tau\simeq 1.477
  M_*$, due to imperfections of our AMR numerics, and sensitivity to
  the choices of $\mu_1$ and $\mu_2$. Hence only the
  collapse stage of the evolution is shown in this figure.}  Although
the resulting dynamics is quite complicated, featuring multiple
scatterings and interferences, there are 3 prominent peaks---marked by
the arrows in Fig. \ref{fig_echos_obt}---that can be identified as
echos. Their temporal period is $\Delta_\tau =1.10\pm0.04$, and on
each echo the logarithm of $|I|$ grows by a comparable amount
$\Delta_r =1.12\pm 0.06$. We note that these values match within the
errorbars, and are in a good agreement with the periods computed in Fig. \ref{fig_echos}.

The echoing is not specific to the curvature invariant $I$, other
metric functions oscillate as well. However, while the echoes of $I$
are signalled by the sharp peaks the fluctuations of metric components
are typically milder, showing up as inflection points (see bottom
panels in Fig. \ref{fig_peaks_obt}).  This makes $I$ a superior
quantity for the purpose of measuring the echoing periods. Although we
mainly discussed variations of $I$ at the location of its global
maximum, we verified that curvature develops echoes in other locations
as well, where, however, the amount of the echos and their amplitude
is generally smaller than around the accumulation locus. Since away
from the accumulation locus the curvature remains bounded in the
critical limit we expect only a finite number of such oscillations.

We conclude this section by briefly discussing the accuracy of our
code. While it was not possible to perform a direct convergence test 
of e.g. the critical amplitude or the scaling exponent, since changes
of the resolution usually required readjustments of the
dissipation, $\eps_{KO}$, and the constraint damping, $\kappa$ and the
gauge parameters $\mu_1, \mu_2$, which alter the
``conditions of the convergence study'', that require all parameters except $h$
to stay fixed.  Nevertheless, as indicated in Table \ref{tab_sigs} the
critical amplitude seems to converge as a function of the resolution, at least in the
equal-$\sig$'s case. In addition, formal numerical convergence tests in
individual, fixed amplitude simulations along with the
demonstration that the Hamiltonian and the momentum constraints are
satisfied during the evolutions were carried out in \cite{ES_2+1+n},
indicating nearly second order convergence, and exponential decay of
the $l_2$-norms of the  constraints at late times.

\begin{figure}[t!]
  \centering 
  \includegraphics[width=9cm]{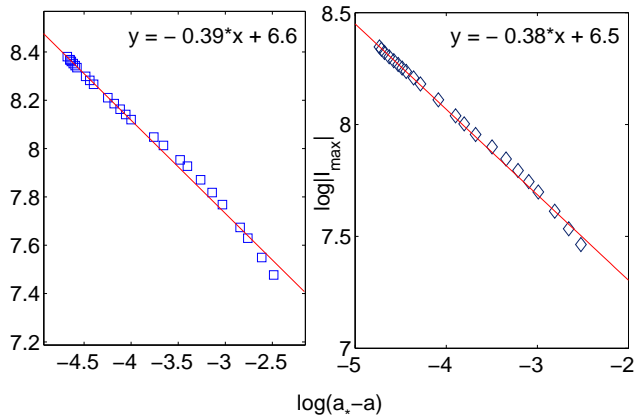}
  \caption[]{The logarithm of $|I_{max}|$ obtained in simulations with
    two distinct sets of the damping and dissipation constants: $\kp =
    1.4, \eps_{KO}=0.75 $ (right panel), and $\kp = 1.5 ,
    \eps_{KO}=0.85$ (left panel). We use same resolution of $h=1/300$,
    and all other equal parameters.  The difference in $\bt$'s in this
    case is less than $3\%$, indicating the quality of our numerics. }
  \label{fig_loglogIa_kp14-15}
\end{figure}

The consistency of the scaling exponents obtained in simulations
with a whole different set of parameters (see
e.g. Figs. \ref{fig_loglogIa_obt0913} and \ref{fig_loglogIa_kp17})
indicates robustness of $\bt$, however the overall accuracy of our code can be
estimated by changing only the numerical parameters.  To this end we
performed simulations with different damping and the Kreiss-Oliger
dissipation constants $\kappa$ and $\eps_{KO}$, and otherwise similar
parameters.  Fig. \ref{fig_loglogIa_kp14-15} shows that $\bt$'s
computed in two sets of simulations defined by $\kp = 1.4,
\eps_{KO}=0.75 $, and $\kp = 1.5 , \eps_{KO}=0.85$ differ by less than
$3\%$. 

\section{Discussion}
\label{sec_discussion}
The ring-shaped accumulation loci that we observe in most evolutions
of the time-symmetric 
Brill-wave initial data indicate that the critical solutions in these
cases are genuinely different from those found by Abrahams \& Evans
\cite{AbrahamsEvans} for the ingoing $l=2$ quasi-linear wave data,
where the maximal curvature occurs at the origin. The black hole
radii and their masses found in slightly super-critical evolutions in
\cite{AbrahamsEvans}  are tend to zero in the critical limit,
signalling so-called type II critical phenomenon, characterized by smooth
transition between dispersion and  black hole formation, see Ref. 
\cite{Gundlach_review}. In our case, however, the radii of the
accumulation loci are finite, as are the apparent
horizons that form in supercritical collapse, engulfing the
accumulation loci. Although, at present, neither our numerics is
capable of finding apparent horizons very close to the threshold,
$a_* \ltsim a\ltsim 1.2 \,a_*$, nor it allows to
trace the evolution of the horizons to their endstate, it 
indicates that our critical solutions include quasi-stationary ring-shaped 
formations of finite size and mass.

The situation with the four and two levels AMR simulations is
somewhat puzzling. While the two level simulations are clearly divergent near
$a=6.9401$, which is determined as the critical amplitude in the
four level runs, comparable in resolution unigrid runs do not
encounter any particular difficulties at this amplitude. We believe
such a behaviour may signal another, different critical solution, which is not
resolved by the lower-dimensional unigrid simulations, and which
destabilizes the less accurate two-level runs. However, whether this is
indeed the case requires further investigation.   

In all cases, we found strong evidence that in subcritical non-rotating axisymmetric
vacuum collapse curvature exhibits a power-law scaling as a function
of parametric distance from the threshold for black hole formation. We
numerically evolved several sets of initial Brill-waves defined by
fixed $\sig_r$ and $\sig_z$, and by a tuneable amplitude, $a$, and
checked that in the limit $a\rightarrow a_*$, $|I_{max}|\propto
(a_*-a)^{-\bt}$ with roughly the same exponent as that computed in
supercritical regime by Abrahams and Evans \cite{AbrahamsEvans}, see
Figs.  \ref{fig_loglogIa_obt0913} and \ref{fig_loglogIa_kp17}.  This
demonstrates that quantities with same length dimensions---such as the black hole mass
in \cite{AbrahamsEvans} and the inverse curvature invariant
$I_{max}^{-1}$ here---scale identically.  We
verified that the exponent is relatively insensitive to coordinate
conditions. Since we find that the scaling occurs around a ring-shaped
accumulation locus, which is different from the point-like one of
\cite{AbrahamsEvans}, 
there is no {\it a priori} reason to expect the exponents in both
cases to match. However, the exponents agree, and this, apparently,
indicates that $\bt\simeq 0.35-0.4$ is truly universal and independent
of the initial data, regardless of what critical solution these data
may lead to. \footnote{In this regard it is interesting to observe that
  the critical exponent originally found by Choptuik in scalar-field collapse, $\beta_{SF}
  \simeq 0.374$, is again comparable to what we find here. While this may be just
  a coincidence, it may, alternatively, point to the genuine role
  of the gravity, rather than matter, in critical behaviour in scalar-field collapse.}

There is evidence that the near critical solutions are periodically
self-similar. Specifically, we observe that in the limit $a
\rightarrow a_*$ the curvature invariant $I$ undergoes increasingly
large number of oscillations, whose period in the proper time is
approximately equal to the rate of variation of the curvature on each
echo $\Delta_\tau \simeq \Delta_r \simeq 1.1$, see
Figs. \ref{fig_peaks}, \ref{fig_echos} and \ref{fig_echos_obt}.  We
note that the echoing periods reported in \cite{AbrahamsEvans},
$\Delta \sim 0.6$, differ from ours, which are roughly twice larger in
magnitude.  However, this is probably not too surprising since our critical
solutions are different from theirs, and besides the period
of any specific quantity will typically depend on the particular
combinations of the metric and derivatives, comprising it (for
instance, the quantity $\pa^2\Psi/\pa\tau^2$ is twice more variable
than $\Psi$).  An independent, if circumstantial, signature of
discrete self-similarity is the distinctive ``wiggle'' of the data
points about the leading power-law scaling of $|I_{max}|$, see
Figs. \ref{fig_loglogIa_obt0913} and \ref{fig_loglogIa_kp17}, since
exactly this kind of behavior is expected in the periodically
self-similar systems \cite{GHP}.

An obvious limitation of the current simulations is their maximal
resolution.  Even though a relatively moderate numerical resolutions
of $h\simeq 1/250-1/1000$ have already provided fruitful insights into
the critical behavior, higher resolutions are needed in order to
compute the scaling and echoing constants more accurately. We expect
that much closer approach to threshold will be required. This should
create a longer span of data, enabling a greater accuracy of linear
fits in the plots like Fig. \ref{fig_loglogIa_obt0913} and
\ref{fig_loglogIa_kp17}, which, in turn, will allow unambiguous
computation of $\bt$ and of the wiggle period. A closer approach
$a\rightarrow a_*$ should also multiply the number of the echoes,
allowing a better estimate on their periods.  Clearly, using numerical
meshes of fixed size is not practical for probing the limit
$a\rightarrow a_*$, rather AMR approach should be used. While we have
already experimented with that, our runs often develop premature
instabilities since in the near critical limit the system tends to be
extremely sensitive to numerical and gauge parameters; for instance,
in the 4-level simulations a slight variation of $\kappa$ by a mere
$1\%$ ruins convergence.
We are
currently improving our code in order to locate the optimal parameter
settings, which will enable us to edge the critical limit, the results
of that study will be reported elsewhere.

\ack The computations were performed on the Damiana cluster of the
AEI.


\section*{References}
\begin{thebibliography}{99}

\bibitem{Choptuik} M.~W.~Choptuik, ``Universality And Scaling In
  Gravitational Collapse Of A Massless Scalar Field,'' Phys.\ Rev.\
  Lett.\ {\bf 70}, 9 (1993).



\bibitem{Gundlach_review} C.~Gundlach and J.~M.~Martin-Garcia,
  ``Critical phenomena in gravitational collapse,'' Living Rev.\ Rel.\
  {\bf 10}, 5 (2007)

\bibitem{AbrahamsEvans} A.~M.~Abrahams and C.~R.~Evans, ``Critical
  behavior and scaling in vacuum axisymmetric gravitational
  collapse,'' Phys.\ Rev.\ Lett.\ {\bf 70}, 2980
  (1993), 
  A.~M.~Abrahams and C.~R.~Evans, ``Universality in axisymmetric
  vacuum collapse,'' Phys.\ Rev.\ D {\bf 49}, 3998 (1994).


\bibitem{Seidel_etal} M.~Alcubierre, G.~Allen, B.~Bruegmann,
  G.~Lanfermann, E.~Seidel, W.~M.~Suen and M.~Tobias,
  Phys.\ Rev.\ D {\bf 61}, 041501 (2000) [arXiv:gr-qc/9904013].
\bibitem{ES_2+1+n} E.~Sorkin, ``An axisymmetric generalized harmonic
  evolution code,'' Phys.\ Rev.\ D {\bf 81}, 084062 (2010)
  [arXiv:0911.2011 [gr-qc]].
\bibitem{Koike_etal} T.~Koike, T.~Hara and S.~Adachi, ``Critical
  behavior in gravitational collapse of radiation fluid: A
  Renormalization group (linear perturbation) analysis,'' Phys.\ Rev.\
  Lett.\ {\bf 74}, 5170 (1995)

\bibitem{Garfinkle_sub} D.~Garfinkle and G.~C.~Duncan, ``Scaling of
  curvature in sub-critical gravitational collapse,'' Phys.\ Rev.\ D
  {\bf 58}, 064024 (1998)


\bibitem{SorkinOren} E.~Sorkin and Y.~Oren, ``On Choptuik's scaling in
  higher dimensions,'' Phys.\ Rev.\ D {\bf 71}, 124005 (2005)
  [arXiv:hep-th/0502034].



\bibitem{Friedrich_GH} H.~Friedrich, ``Hyperbolic Reductions For
  Einstein's Equations,'' Class.\ Quant.\ Grav.\ {\bf 13}, 1451
  (1996).

\bibitem{Garfinkle_GH} D.~Garfinkle, ``Harmonic coordinate method for
  simulating generic singularities,'' Phys.\ Rev.\ D {\bf 65}, 044029
  (2002)

\bibitem{Pretorius_GH} F.~Pretorius, ``Numerical Relativity Using a
  Generalized Harmonic Decomposition,'' Class.\ Quant.\ Grav.\ {\bf
    22}, 425 (2005)

\bibitem{DWG} L.~Lindblom and B.~Szilagyi, ``An Improved Gauge Driver
  for the GH Einstein System,'' Phys.\ Rev.\ D {\bf 80}, 084019 (2009)
  M.~W.~Choptuik and F.~Pretorius, ``Ultra Relativistic Particle
  Collisions,'' Phys.\ Rev.\ Lett.\ {\bf 104}, 111101 (2010)


\bibitem{Brill} D.~R.~Brill, ``On the positive definite mass of the
  Bondi-Weber-Wheeler time-symmetric gravitational waves,'' Annals
  Phys.\ {\bf 7}, 466 (1959).

\bibitem{Lindblom_etal1} L.~Lindblom, M.~A.~Scheel, L.~E.~Kidder,
  R.~Owen and O.~Rinne, ``A New Generalized Harmonic Evolution
  System,'' Class.\ Quant.\ Grav.\ {\bf 23}, S447 (2006)


\bibitem{Gundlach_CD} C.~Gundlach, J.~M.~Martin-Garcia, G.~Calabrese
  and I.~Hinder, ``Constraint damping in the Z4 formulation and
  harmonic gauge,'' Class.\ Quant.\ Grav.\ {\bf 22}, 3767 (2005)
  [arXiv:gr-qc/0504114].


\bibitem{Pretorius_BH1} F.~Pretorius, ``Simulation of binary black
  hole spacetimes with a harmonic evolution scheme,'' Class.\ Quant.\
  Grav.\ {\bf 23}, S529 (2006) [arXiv:gr-qc/0602115].


\bibitem{SorkinChoptuik} E.~Sorkin and M.~W.~Choptuik, ``Generalized
  harmonic formulation in spherical symmetry,'' Gen.\ Rel.\ Grav.\
  {\bf 42}, 1239 (2010) [arXiv:0908.2500 [gr-qc]].

\bibitem{Teukolsky_BW} A.~M.~Abrahams, K.~R.~Heiderich, S.~L.~Shapiro
  and S.~A.~Teukolsky, ``Vacuum Initial Data, Singularities, And
  Cosmic Censorship,'' Phys.\ Rev.\ D {\bf 46}, 2452 (1992).

\bibitem{Garfinkle_BW} D.~Garfinkle and G.~C.~Duncan, ``Numerical
  evolution of Brill waves,'' Phys.\ Rev.\ D {\bf 63}, 044011 (2001)
  [arXiv:gr-qc/0006073].

\bibitem{pamr} Parallel Adaptive Mesh Refinement (PAMR) and Adaptive
  Mesh Refinement Driver (AMRD), \texttt{
    http://laplace.phas.ubc.ca/Group/Software.html.}


\bibitem{HamadeStewart} R.~S.~Hamade and J.~M.~Stewart, ``The
  Spherically symmetric collapse of a massless scalar field,'' Class.\
  Quant.\ Grav.\ {\bf 13}, 497 (1996) [arXiv:gr-qc/9506044].



\bibitem{GHP} C.~Gundlach, ``Understanding critical collapse of a
  scalar field,'' Phys.\ Rev.\ D {\bf 55}, 695 (1997)
  S.~Hod and T.~Piran, ``Fine-structure of Choptuik's mass-scaling
  relation,'' Phys.\ Rev.\ D {\bf 55}, 440 (1997)



\end{thebibliography}
\end{document}